\documentclass[twocolumn,letterpaper,aps,prc,superscriptaddress,nofootinbib,floatfix]{revtex4}

\usepackage{xspace}
\usepackage{graphicx}
\usepackage[hypertex]{hyperref}
\usepackage{multirow}
\usepackage{natbib}

\newcommand{\raa}{\mbox{$R_{\rm AA}$}\xspace}
\newcommand{\taa}{\mbox{$T_{\rm AA}$}\xspace}
\newcommand{\rcp}{\mbox{$R_{\rm CP}$}\xspace}
\newcommand{\geant}{\mbox{\sc geant}\xspace}
\newcommand{\pythia}{\mbox{\sc pythia}\xspace}
\newcommand{\ncoll}{\mbox{$N_{\rm coll}$}\xspace}
\newcommand{\npart}{\mbox{$N_{\rm part}$}\xspace}
\newcommand{\auau}{\mbox{Au$+$Au}\xspace}
\newcommand{\dau}{\mbox{$d$+Au}\xspace}
\newcommand{\pbpb}{\mbox{Pb+Pb}\xspace}

\newcommand{\pt}{\mbox{$p_T$}\xspace}
\newcommand{\jpsi}{\mbox{$J/\psi$}\xspace}
\newcommand{\jpsis}{\mbox{$J/\psi$s}\xspace}
\newcommand{\psip}{\mbox{$\psi^{\prime}$}\xspace}

\newcommand{\pp}{\mbox{$p$+$p$}\xspace}

\newcommand{\snn}{\mbox{$\sqrt{s_{_{NN}}}$}\xspace}

\begin{document}


\title{$J/\psi$ suppression at forward rapidity in \auau collisions at 
$\sqrt{s_{_{NN}}}$=39 and 62.4 GeV}

\newcommand{\abilene}{Abilene Christian University, Abilene, Texas 79699, USA}
\newcommand{\banaras}{Department of Physics, Banaras Hindu University, Varanasi 221005, India}
\newcommand{\barc}{Bhabha Atomic Research Centre, Bombay 400 085, India}
\newcommand{\baruch}{Baruch College, City University of New York, New York, New York, 10010 USA}
\newcommand{\bnlcoll}{Collider-Accelerator Department, Brookhaven National Laboratory, Upton, New York 11973-5000, USA}
\newcommand{\bnlphys}{Physics Department, Brookhaven National Laboratory, Upton, New York 11973-5000, USA}
\newcommand{\caucr}{University of California - Riverside, Riverside, California 92521, USA}
\newcommand{\charlesczech}{Charles University, Ovocn\'{y} trh 5, Praha 1, 116 36, Prague, Czech Republic}
\newcommand{\chonbuk}{Chonbuk National University, Jeonju, 561-756, Korea}
\newcommand{\cns}{Center for Nuclear Study, Graduate School of Science, University of Tokyo, 7-3-1 Hongo, Bunkyo, Tokyo 113-0033, Japan}
\newcommand{\colorado}{University of Colorado, Boulder, Colorado 80309, USA}
\newcommand{\columbia}{Columbia University, New York, New York 10027 and Nevis Laboratories, Irvington, New York 10533, USA}
\newcommand{\czechtech}{Czech Technical University, Zikova 4, 166 36 Prague 6, Czech Republic}
\newcommand{\dapnia}{Dapnia, CEA Saclay, F-91191, Gif-sur-Yvette, France}
\newcommand{\debrecen}{Debrecen University, H-4010 Debrecen, Egyetem t{\'e}r 1, Hungary}
\newcommand{\elte}{ELTE, E{\"o}tv{\"o}s Lor{\'a}nd University, H - 1117 Budapest, P{\'a}zm{\'a}ny P. s. 1/A, Hungary}
\newcommand{\ewha}{Ewha Womans University, Seoul 120-750, Korea}
\newcommand{\fsu}{Florida State University, Tallahassee, Florida 32306, USA}
\newcommand{\gsu}{Georgia State University, Atlanta, Georgia 30303, USA}
\newcommand{\hanyang}{Hanyang University, Seoul 133-792, Korea}
\newcommand{\hiroshima}{Hiroshima University, Kagamiyama, Higashi-Hiroshima 739-8526, Japan}
\newcommand{\ihepprot}{IHEP Protvino, State Research Center of Russian Federation, Institute for High Energy Physics, Protvino, 142281, Russia}
\newcommand{\illuiuc}{University of Illinois at Urbana-Champaign, Urbana, Illinois 61801, USA}
\newcommand{\inrras}{Institute for Nuclear Research of the Russian Academy of Sciences, prospekt 60-letiya Oktyabrya 7a, Moscow 117312, Russia}
\newcommand{\instpasczech}{Institute of Physics, Academy of Sciences of the Czech Republic, Na Slovance 2, 182 21 Prague 8, Czech Republic}
\newcommand{\isu}{Iowa State University, Ames, Iowa 50011, USA}
\newcommand{\jyvaskyla}{Helsinki Institute of Physics and University of Jyv{\"a}skyl{\"a}, P.O.Box 35, FI-40014 Jyv{\"a}skyl{\"a}, Finland}
\newcommand{\kek}{KEK, High Energy Accelerator Research Organization, Tsukuba, Ibaraki 305-0801, Japan}
\newcommand{\korea}{Korea University, Seoul, 136-701, Korea}
\newcommand{\kurchatov}{Russian Research Center ``Kurchatov Institute", Moscow, 123098 Russia}
\newcommand{\kyoto}{Kyoto University, Kyoto 606-8502, Japan}
\newcommand{\labllr}{Laboratoire Leprince-Ringuet, Ecole Polytechnique, CNRS-IN2P3, Route de Saclay, F-91128, Palaiseau, France}
\newcommand{\lawllnl}{Lawrence Livermore National Laboratory, Livermore, California 94550, USA}
\newcommand{\losalamos}{Los Alamos National Laboratory, Los Alamos, New Mexico 87545, USA}
\newcommand{\lpc}{LPC, Universit{\'e} Blaise Pascal, CNRS-IN2P3, Clermont-Fd, 63177 Aubiere Cedex, France}
\newcommand{\lund}{Department of Physics, Lund University, Box 118, SE-221 00 Lund, Sweden}
\newcommand{\maryland}{University of Maryland, College Park, Maryland 20742, USA}
\newcommand{\mass}{Department of Physics, University of Massachusetts, Amherst, Massachusetts 01003-9337, USA }
\newcommand{\muhlenberg}{Muhlenberg College, Allentown, Pennsylvania 18104-5586, USA}
\newcommand{\myongji}{Myongji University, Yongin, Kyonggido 449-728, Korea}
\newcommand{\nagasaki}{Nagasaki Institute of Applied Science, Nagasaki-shi, Nagasaki 851-0193, Japan}
\newcommand{\newmex}{University of New Mexico, Albuquerque, New Mexico 87131, USA }
\newcommand{\nmsu}{New Mexico State University, Las Cruces, New Mexico 88003, USA}
\newcommand{\ohio}{Department of Physics and Astronomy, Ohio University, Athens, Ohio 45701, USA}
\newcommand{\ornl}{Oak Ridge National Laboratory, Oak Ridge, Tennessee 37831, USA}
\newcommand{\orsay}{IPN-Orsay, Universite Paris Sud, CNRS-IN2P3, BP1, F-91406, Orsay, France}
\newcommand{\pnpi}{PNPI, Petersburg Nuclear Physics Institute, Gatchina, Leningrad region, 188300, Russia}
\newcommand{\riken}{RIKEN Nishina Center for Accelerator-Based Science, Wako, Saitama 351-0198, Japan}
\newcommand{\rikjrbrc}{RIKEN BNL Research Center, Brookhaven National Laboratory, Upton, New York 11973-5000, USA}
\newcommand{\rikkyo}{Physics Department, Rikkyo University, 3-34-1 Nishi-Ikebukuro, Toshima, Tokyo 171-8501, Japan}
\newcommand{\saispbstu}{Saint Petersburg State Polytechnic University, St. Petersburg, 195251 Russia}
\newcommand{\saopaulo}{Universidade de S{\~a}o Paulo, Instituto de F\'{\i}sica, Caixa Postal 66318, S{\~a}o Paulo CEP05315-970, Brazil}
\newcommand{\seoulnat}{Department of Physics and Astronomy, Seoul National University, Seoul, Korea}
\newcommand{\stonybrkc}{Chemistry Department, Stony Brook University, SUNY, Stony Brook, New York 11794-3400, USA}
\newcommand{\stonycrkp}{Department of Physics and Astronomy, Stony Brook University, SUNY, Stony Brook, New York 11794-3400, USA}
\newcommand{\tenn}{University of Tennessee, Knoxville, Tennessee 37996, USA}
\newcommand{\titech}{Department of Physics, Tokyo Institute of Technology, Oh-okayama, Meguro, Tokyo 152-8551, Japan}
\newcommand{\tsukuba}{Institute of Physics, University of Tsukuba, Tsukuba, Ibaraki 305, Japan}
\newcommand{\vandy}{Vanderbilt University, Nashville, Tennessee 37235, USA}
\newcommand{\weizmann}{Weizmann Institute, Rehovot 76100, Israel}
\newcommand{\wigner}{Institute for Particle and Nuclear Physics, Wigner Research Centre for Physics, Hungarian Academy of Sciences (Wigner RCP, RMKI) H-1525 Budapest 114, POBox 49, Budapest, Hungary}
\newcommand{\yonsei}{Yonsei University, IPAP, Seoul 120-749, Korea}
\affiliation{\abilene}
\affiliation{\banaras}
\affiliation{\barc}
\affiliation{\baruch}
\affiliation{\bnlcoll}
\affiliation{\bnlphys}
\affiliation{\caucr}
\affiliation{\charlesczech}
\affiliation{\chonbuk}
\affiliation{\cns}
\affiliation{\colorado}
\affiliation{\columbia}
\affiliation{\czechtech}
\affiliation{\dapnia}
\affiliation{\debrecen}
\affiliation{\elte}
\affiliation{\ewha}
\affiliation{\fsu}
\affiliation{\gsu}
\affiliation{\hanyang}
\affiliation{\hiroshima}
\affiliation{\ihepprot}
\affiliation{\illuiuc}
\affiliation{\inrras}
\affiliation{\instpasczech}
\affiliation{\isu}
\affiliation{\jyvaskyla}
\affiliation{\kek}
\affiliation{\korea}
\affiliation{\kurchatov}
\affiliation{\kyoto}
\affiliation{\labllr}
\affiliation{\lawllnl}
\affiliation{\losalamos}
\affiliation{\lpc}
\affiliation{\lund}
\affiliation{\maryland}
\affiliation{\mass}
\affiliation{\muhlenberg}
\affiliation{\myongji}
\affiliation{\nagasaki}
\affiliation{\newmex}
\affiliation{\nmsu}
\affiliation{\ohio}
\affiliation{\ornl}
\affiliation{\orsay}
\affiliation{\pnpi}
\affiliation{\riken}
\affiliation{\rikjrbrc}
\affiliation{\rikkyo}
\affiliation{\saispbstu}
\affiliation{\saopaulo}
\affiliation{\seoulnat}
\affiliation{\stonybrkc}
\affiliation{\stonycrkp}
\affiliation{\tenn}
\affiliation{\titech}
\affiliation{\tsukuba}
\affiliation{\vandy}
\affiliation{\weizmann}
\affiliation{\wigner}
\affiliation{\yonsei}
\author{A.~Adare} \affiliation{\colorado}
\author{C.~Aidala} \affiliation{\losalamos}
\author{N.N.~Ajitanand} \affiliation{\stonybrkc}
\author{Y.~Akiba} \affiliation{\riken} \affiliation{\rikjrbrc}
\author{R.~Akimoto} \affiliation{\cns}
\author{H.~Al-Ta'ani} \affiliation{\nmsu}
\author{J.~Alexander} \affiliation{\stonybrkc}
\author{A.~Angerami} \affiliation{\columbia}
\author{K.~Aoki} \affiliation{\riken}
\author{N.~Apadula} \affiliation{\stonycrkp}
% ...
\collaboration{PHENIX Collaboration} \noaffiliation


\begin{abstract}

We present measurements of the $J/\psi$ invariant yields in 
$\sqrt{s_{_{NN}}}$=39 and 62.4 GeV \auau collisions at forward rapidity 
($1.2<|y|<2.2$). Invariant yields are presented as a function of both 
collision centrality and transverse momentum. Nuclear modifications are 
obtained for central relative to peripheral Au$+$Au collisions 
($R_{\rm CP}$) and for various centrality selections in \auau relative 
to scaled $p$$+$$p$ cross sections obtained from other measurements 
($R_{\rm AA}$).  The observed suppression patterns at 39 and 62.4 GeV are 
quite similar to those previously measured at 200 GeV.  This similar 
suppression presents a challenge to theoretical models that contain various 
competing mechanisms with different energy dependencies, some of which cause 
suppression and others enhancement.

\end{abstract}

\date{\today}

\maketitle

Heavy quarkonia are bound states of charm-anticharm or bottom-antibottom 
quarks. It was proposed over 25 years ago that these states would be color 
screened in a quark-gluon plasma (QGP), thus suppressing their final 
yields in relativistic heavy ion collisions~\cite{Matsui:1986}.  The NA50 
experiment at the CERN-SPS measured a significant suppression of \jpsi and 
\psip in \pbpb collisions at \snn = 17.2 GeV, which was interpreted as 
indicating the onset of quark-gluon plasma 
formation~\cite{Abreu:2000p3069}.  However, measurements by the PHENIX 
experiment at the Relativistic Heavy Ion Collider (RHIC) indicated a 
similar level of nuclear suppression at midrapidity in \auau collisions at 
\snn= 200 GeV~\cite{Adare:2006ns}.  Additional PHENIX results also 
indicated a larger suppression at forward rapidity $1.2 < |y| < 2.2$ 
compared with midrapidity, despite the expectation of a higher energy 
density and temperature for the medium at midrapidity. Perhaps more 
surprising is the comparison of the recent higher statistics PHENIX 
forward rapidity \jpsi suppression~\cite{PhysRevC.84.054912} and the ALICE 
experiment measurement in \pbpb at 2.76 TeV~\cite{Abelev:2012rv} at the 
Large Hadron Collider (LHC). These results indicate significantly less 
suppression for the most central \pbpb events at the LHC compared with 
\auau events at RHIC. Results at RHIC and the LHC at larger transverse 
momentum ($p_T >$ 4 
GeV/$c$)~\cite{Aad:2010aa,Chatrchyan:2012np,Tang:2011kr} suggest the 
opposite, with more suppression at the LHC compared to RHIC.

These measurements contradict an interpretation based solely on color 
screening, and require the influence of other physics. There is an 
additional class of effects referred to as ``cold nuclear matter'' (CNM) 
effects that are not due to the creation of a hot medium and thus can be 
probed via $p(d)$+$A$ collisions. These CNM effects include the 
modification of the initial incoming flux of quarks and gluons in the 
nucleus as described by nuclear-modified parton distribution functions 
(nPDFs)~\cite{Eskola:2009uj}, breakup of the \jpsi precursor 
$c\overline{c}$ state while traversing the nucleus, and initial-state 
parton energy loss. CNM effects have been studied in detail in $p$+$A$ 
collisions at \snn = 17-42 GeV~\cite{Abreu:1998ee, Alessandro:2003pi, 
Alessandro:2003pc, Alessandro:2006jt, Arnaldi:2010ky, 
FNAL:NuSea,Abt:2008ya}, in \dau collisions at \snn = 200 GeV by the 
PHENIX experiment~\cite{Adare:2010fn,Adare:2012qf,Nagle:2010ix}, and 
$p$+$A$ results from the LHC are anxiously awaited. In addition, there may 
be effects in the QGP other than color screening.  These include the 
possible coalescence of originally uncorrelated $c$ and $\overline{c}$ 
quarks or the recombination of breakup $c$ and $\overline{c}$ pairs 
resulting in a competing enhancement effect (see for example 
Refs.~\cite{Thews:2006p3770,Zhou:2009vz}). This coalescence effect is 
expected to grow as the density of $c$ and $\overline{c}$ increases. A 
recent review of many of these phenomena is given in 
Ref.~\cite{Brambilla:2010cs}.

All of this highlights the importance of measuring \jpsi and other excited 
quarkonia states over a broad range in \snn; thus varying not only the 
temperature of the medium, but also the $c$ and $\overline{c}$ production 
and the cold nuclear matter effects.  In this paper, the PHENIX 
collaboration presents first measurements of invariant yields and 
suppression for \jpsi at forward rapidity $1.2 < |y| < 2.2$ in \auau 
collisions at \snn = 39 and 62.4 GeV.

\section{Data analysis}

The PHENIX experiment collected data in 2010 for \auau collisions at \snn 
= 39 and 62.4 GeV as part of the RHIC Beam Energy Scan program.  After 
good run selection cuts, the data set includes $2.0 \times 10^{8}$ events 
at 39 GeV and $5.5 \times 10^{8}$ events at 62.4 GeV. The PHENIX 
experiment is described in detail in Ref.~\cite{Adcox:2003p2584}.  The 
\jpsi measurement at forward rapidity is made via the dimuon decay channel 
with two forward angle muon spectrometers, as detailed in 
Ref.~\cite{Aronson:2003p2579}.  The muon spectrometers have acceptance 
over the range $1.2 < |\eta| < 2.2$ and over the full azimuth.  The two 
spectrometers comprise an initial hadronic absorber followed by three sets 
of cathode strip chambers which are inside a magnetic field, referred to 
as the Muon Tracker (MuTr), and then five planes of Iarocci tubes 
interleaved with steel absorber plates, referred to as the Muon Identifier 
(MuID).  Muon candidates are found by reconstructing tracks through the 
magnetic field in the MuTr and matching them to MuID tracks that penetrate 
through to the last MuID plane.

In \auau collisions at \snn = 39 and 62.4 GeV, the events are selected 
with a minimum bias (MB) trigger utilizing the Beam-Beam Counter (BBC). 
The BBC comprises two arrays of 64 quartz \v{C}erenkov counters covering 
pseudorapidity $3.0 < |\eta| < 3.9$.  The MB trigger requires at least two 
hits in each of the BBC arrays and a reconstructed collision $z$-vertex of 
$|z| < 30$ cm, where $z = 0$ is the center of the detector.  The BBC total 
charge is used as a measure of the collision centrality (the impact 
parameter of the \auau collision is monotonically related to the average 
total charged particle multiplicity).  Following the procedure used for 
\auau at \snn = 200 GeV, for each centrality selection the average number 
of nucleon participants ($\langle \npart \rangle$) and the average number 
of binary collisions ($\langle \ncoll \rangle$) are estimated using a 
Glauber model of the collision~\cite{Miller:2007p723} and a negative 
binomial parametrization of the charged particles per pair of 
participating nucleons. The total fraction of the \auau inelastic cross 
section measured by the MB trigger is determined to be $85.7 \pm 2.0$\% 
and $85.9 \pm 2.0$\% at 39 and 62.4 GeV, respectively.  The minimum bias 
sample is divided into exclusive centrality bins that are categorized via 
the Glauber model comparison to the BBC charge distribution as given in 
Table~\ref{tab:glauber}.  Note that the centrality selections used here 
are wider than in previous analyses for \auau collisions at \snn = 200 GeV 
due to the smaller statistical sample of \jpsis.


\begin{table}[!ht]

\caption{\label{tab:glauber} 
Mean $\npart$ and $\ncoll$ values and systematic uncertainties in each 
centrality bin for \auau at 39 and 62.4 GeV.
}
\begin{ruledtabular}\begin{tabular}{cccc}
$\sqrt{s}(GeV)$ & Cent(\%) & $\langle\npart\rangle$ & $\langle\ncoll\rangle$ \\
\hline
\multirow{2}{*}{39} & 0-40 & 204.4 $\pm$ 4.4 & 444.8 $\pm$ 50.3 \\
& 40-86 & 34.1 $\pm$ 1.6 & 43.5 $\pm$ 3.7 \\ 
\\
\multirow{4}{*}{62.4} & 0-20 & 274.8 $\pm$ 3.8 & 689.9 $\pm$ 78.9 \\
& 20-40 & 138.7 $\pm$ 4.7 & 270.5 $\pm$ 27.5 \\
& 40-60 & 59.7 $\pm$ 3.9 & 85.7 $\pm$ 9.1 \\
& 60-86 & 14.7 $\pm$ 1.2 & 14.3 $\pm$ 1.7 \\
\end{tabular} \end{ruledtabular}
\end{table}

\begin{figure}
\centering
\includegraphics[width=0.90\linewidth]{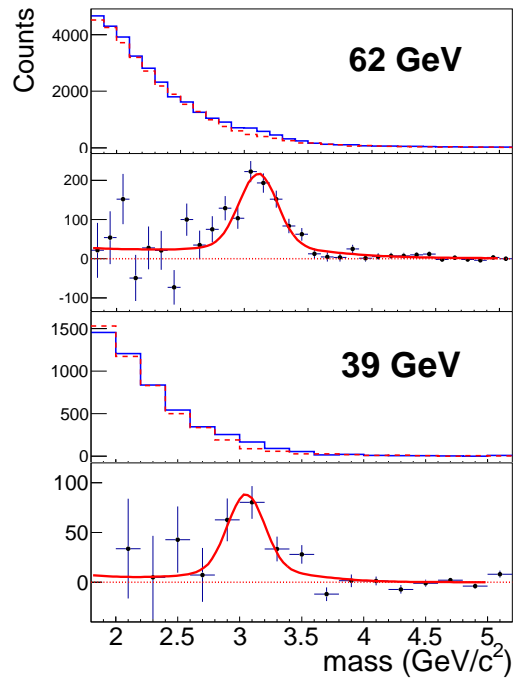}
\caption{\label{fig:massplots} (color online) 
The unlike sign invariant mass distribution (blue) for all \auau 
centralities and both muon arms is shown along with the uncorrelated 
background calculation from mixed event pairs (red) for 62 GeV (top two 
panels) and 39 GeV (bottom two panels). The lower panel in each pair shows 
the subtracted distribution and a fit to the data.
}
\end{figure}

For each centrality selection and beam energy, we extract the number of 
\jpsi counts following a method identical to that used in 
Ref.~\cite{PhysRevC.84.054912}.  All unlike-sign muon candidates are 
paired to calculate the invariant mass distribution.  Underneath the \jpsi 
signal are continuum background counts both from uncorrelated tracks and 
from correlated physical backgrounds such as open charm decay (e.g. 
$D\overline{D}$ where both decay semi-leptonically to muons), open bottom 
decay, and the Drell-Yan process.  First the uncorrelated background is 
calculated via an event mixing method with pairs from different \auau 
events with the same centrality and $z$-vertex. This background is then 
normalized using a comparison of real-event and mixed-event like-sign 
pairs.  After subtraction of the uncorrelated background, we fit to the 
remaining correlated dimuon spectrum with an acceptance-modulated \jpsi 
line shape (determined from a full \geant~\cite{geant3} simulation of the 
PHENIX detector) and an exponential folded with the acceptance to model 
the remaining correlated physics background.  Utilizing different 
assumptions about the line shape, different uncorrelated background 
normalizations, and different invariant mass ranges for the fit (as 
detailed in Ref.~\cite{PhysRevC.84.054912}), we determine the systematic 
uncertainty on the extracted \jpsi signal counts.  The total \jpsi sample 
corresponding to all centralities is approximately 170 counts at \snn = 39 
GeV and approximately 1060 counts at \snn = 62.4 GeV.  The invariant mass 
distribution of unlike-sign pairs, mixed-event pairs, and the subtracted 
distributions are shown in Figure~\ref{fig:massplots}. The signal 
extraction procedure is quite robust and the systematic uncertainty is of 
order 2-10\%.

The \jpsi invariant yield is expressed as:

\begin{equation}
B_{\mu\mu} \frac{d^3N}{d\pt^2 dy} = \frac{1}{2\pi\pt{\scriptstyle\Delta}\pt{\scriptstyle\Delta} y}\frac{N_{J/\psi}}{A\epsilon\ N_{EVT}}
\end{equation}

\noindent where $B_{\mu\mu}$ is the branching fraction of $\jpsi$ to 
muons, $N_{\jpsi}$ is the number of measured $\jpsi$s, $N_{EVT}$ is the 
number of events in the relevant \auau{} centrality selection, $A\epsilon$ 
is the detector geometric acceptance times efficiency, and 
${\scriptstyle\Delta}\pt$ and ${\scriptstyle\Delta}y$ are the bin width in 
$\pt$ and $y$, respectively.  For the $\pt$-integrated bins, we similarly 
calculate $B_{\mu\mu}dN/dy = N_{J/\psi}/(A\epsilon N_{EVT} 
{\scriptstyle\Delta} y)$. We evaluate the acceptance and reconstruction 
efficiency by running \pythia-generated~\cite{pythia} \jpsis through the 
{\sc geant} simulation of the PHENIX detector and then embedding these 
simulated hits into real \auau data events.  These simulated events are 
then reconstructed using identical code to that used in the real data 
analysis, and the overall acceptance and efficiency ($A\epsilon$) is 
determined for each \auau centrality selection. In measurements at higher 
energy~\cite{PhysRevC.84.054912}, where the multiplicity is larger, there 
are large drops in the efficiency for more central collisions; but for 
these lower energies, with their lower multiplicity, there is no 
significant loss of efficiency for central collisions. There is an 
additional check on the efficiency of each MuTr and MuID plane that is 
determined via a data-driven method.  The invariant yields are calculated 
separately for each of the two muon spectrometers and then a weighted 
average taken.  These results agree within uncertainties in all cases.

Two categories of systematic uncertainties on the invariant yields are 
shown in Table~\ref{tab:syserr}: type A are point-to-point uncorrelated, 
and type B are correlated (or anti-correlated) point-to-point. The 
uncertainties listed in order are from uncertainties on the \jpsi 
extracted yield as described above, the detector acceptance, the 
acceptance and efficiency over the rapidity range $1.2 < |y| < 2.2$ from 
the assumed {\sc pythia} input distribution, the absolute check on the 
MuTr and MuID hit efficiencies, and the matching of dead areas in the real 
data and {\sc geant} Monte Carlo (MC) simulation.

\begin{table}[!ht]
\caption{\label{tab:syserr} Systematic uncertainties}
\begin{ruledtabular}\begin{tabular}{ccc}
Description & Contribution & Type \\
\hline
Yield extraction & 2-10\% & A \\
Detector acceptance & 5\% & B \\
Input y,$\pt$ distribution & 4\% & B \\
MuTr efficiency & 2\% & B \\
MuID efficiency & 4\% & B \\
DATA and MC mismatch & 4\% & B \\
\end{tabular}\end{ruledtabular}
\end{table}

\section{Results}

\begin{figure}
\centering
\includegraphics[width=1.0\linewidth]{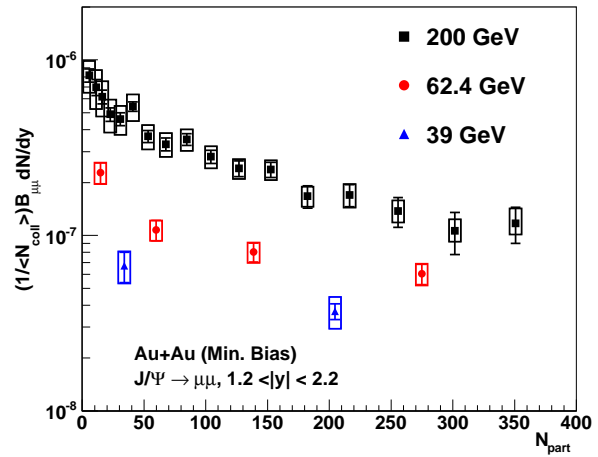}
\caption{\label{fig:invncoll} (color online) 
\jpsi invariant yields (scaled by $1/\langle\ncoll\rangle$) are shown for 
\auau collisions at 39, 62.4, and 200 GeV as a function of the number of 
participating nucleons. The solid error bars represent the uncorrelated 
point-to-point uncertainties (quadrature sum of statistical and type A); 
and the boxes represent the correlated (type B) systematic uncertainties.
}
\end{figure}

\begin{figure}
\centering
\includegraphics[width=1.0\linewidth]{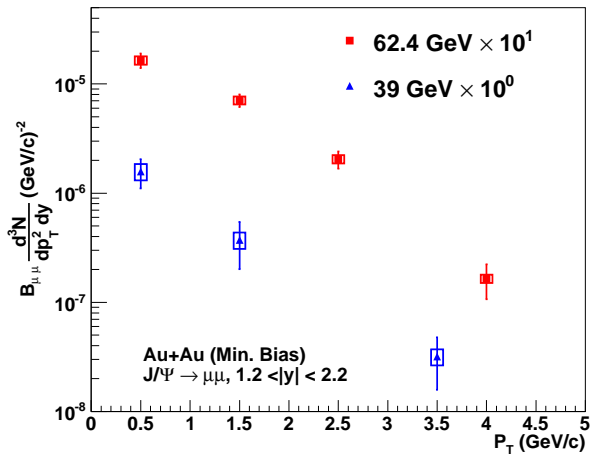}
\caption{\label{fig:invpt} (color online) 
\jpsi invariant yields in minimum bias \auau collisions at 39 and 62.4 GeV 
as a function of transverse momentum. The solid error bars are the 
quadrature sum of the statistical and type A systematic uncertainties, and 
the boxes represent the correlated (type B) systematic uncertainties.
}
\end{figure}

Figure~\ref{fig:invncoll} shows the final calculated \jpsi invariant yield 
integrated over all \pt in \snn = 39 and 62.4 GeV \auau collisions as a 
function of centrality, categorized by the average number of participants 
$\langle \npart \rangle$.  The yields have been rescaled by $1/\langle 
\ncoll \rangle$.  For comparison, the previously published \jpsi invariant 
yields in the same rapidity range $1.2 < |y| < 2.2$ from \snn = 200 GeV 
\auau collisions are also shown~\cite{PhysRevC.84.054912}.  The vertical 
error bars are the quadrature sum of the statistical and type A systematic 
uncertainties, and the boxes represent the type B uncertainties. As 
expected, the \jpsi yield is larger in \auau collisions at larger 
center-of-mass energy.  In addition, the yield per binary collision is 
decreasing with $\langle \npart \rangle$ at all three energies, indicating 
increasing nuclear suppression for more central collisions. 
Figure~\ref{fig:invpt} shows the invariant yield as a function of \pt, 
plotted at the center of each \pt bin, for \snn = 39 and 62.4 GeV \auau 
collisions.

The nuclear modification of \jpsi yields can be categorized in different 
ways.  Because the PHENIX experiment has not yet measured the \pp 
reference baselines at \snn = 39 or 62.4 GeV, we first discuss the \jpsi 
\rcp, the nuclear modification of central relative to peripheral classes 
of events as defined below:

\begin{eqnarray}
 \rcp = \frac {\frac {dN_{\rm AuAu}/ dy}{\langle\ncoll\rangle} (central)}  
 {\frac {dN_{\rm AuAu}/dy} {\langle\ncoll\rangle} (peripheral)} 
\end{eqnarray}

\begin{figure}
\centering
\includegraphics[width=1.0\linewidth]{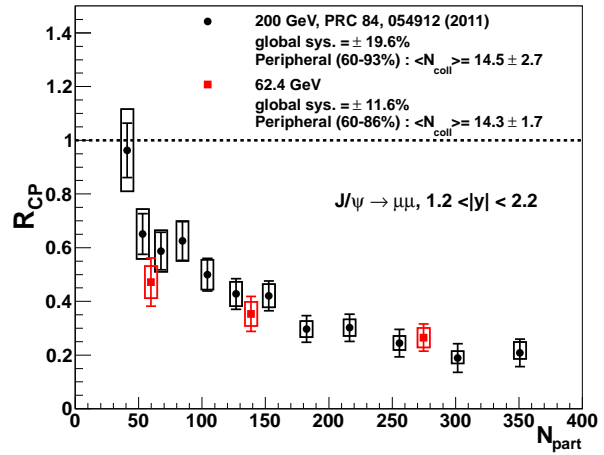}
\caption{\label{fig:rcp62} (color online) 
\jpsi \rcp for 0-20\%, 20-40\%, and 40-60\% (central) relative to 60-86\% 
(peripheral) \auau collisions at 62.4 GeV.  For comparison, \rcp results 
from \auau collisions at 200 GeV are shown with a peripheral bin of 
60-93\%, where the $\langle \ncoll \rangle$ value is a close match. The 
solid error bars are the quadrature sum of the statistical and type A 
systematic uncertainties, and the boxes represent the correlated (type B) 
systematic uncertainties.
}
\end{figure}

\noindent The \rcp values are shown in Figure~\ref{fig:rcp62} and in 
Table~\ref{tab:39_62_rcp} for \auau at 62.4 GeV. Note that the peripheral 
bin selection for \auau at 62.4 GeV is 60-86\% centrality with a 
corresponding $\langle \ncoll \rangle = 14.3 \pm 1.7$. Many uncertainties 
in the invariant yields cancel for \rcp and the dominant uncertainties are 
from the normalization with respect to the peripheral bin including the 
uncertainties in the $\langle \ncoll \rangle$ values for each bin. There 
is an additional type C global systematic from the uncertainty in the 
peripheral $\langle \ncoll \rangle$ value listed in the figure legend and 
in Table~\ref{tab:39_62_rcp}; the other systematic uncertainties are 
included in the boxes on each data point. For comparison, we show the 
published \auau results at 200 GeV~\cite{PhysRevC.84.054912} where the 
peripheral selection is 60-93\%, with a quite comparable $\langle \ncoll 
\rangle = 14.5 \pm 2.7$.  Within uncertainties, the centrality-dependent 
nuclear modification from peripheral to central collisions at the two 
energies are the same.

\begin{table}[!ht]
\caption{\label{tab:39_62_rcp} 
PHENIX 39 and 62.4 GeV \jpsi $R_{CP}$ vs Centrality
with statistical uncertainties and Type A, B, and C systematics.}
\begin{ruledtabular}
\begin{tabular}{ccccccc}
$\sqrt{s}(GeV)$ & Cent(\%) & $R_{CP}$ & Stat & Type A & Type B & Type C \\
\hline
39       & 0-40   & 0.554 & 0.112 & 0.028 & 0.138 & 0.047 \\
\\
62.4     & 0-20   & 0.266 & 0.050 & 0.005 & 0.036 & 0.031 \\
         & 20-40  & 0.353 & 0.064 & 0.008 & 0.045 & 0.041 \\
         & 40-60  & 0.471 & 0.089 & 0.013 & 0.060 & 0.055 \\
\end{tabular}\end{ruledtabular}
\end{table}

\begin{figure}
\centering
\includegraphics[width=1.0\linewidth]{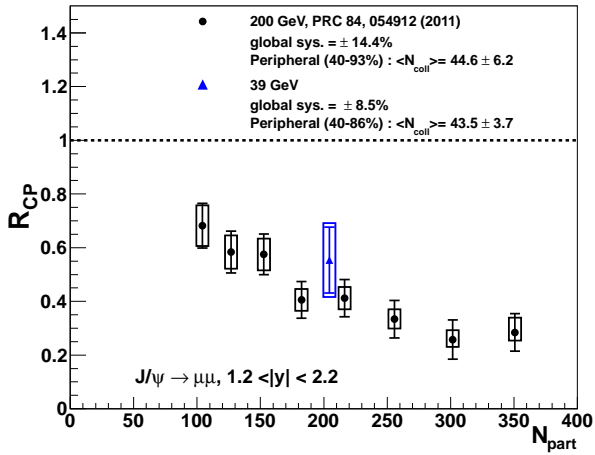}
\caption{\label{fig:rcp39} (color online) 
\jpsi \rcp for 0-40\% (central) relative to 40-87\% (peripheral) \auau 
collisions at 39 GeV. For comparison, \rcp results from \auau collisions 
at 200 GeV are shown with a peripheral bin of 40-93\%, where the $\langle 
\ncoll \rangle$ value is a close match. The solid error bars are the 
quadrature sum of the statistical and type A systematic uncertainties, and 
the boxes represent the correlated (type B) systematic uncertainties.
}
\end{figure}

For the \auau results at 39 GeV, the statistics do not allow any 
centrality dependence of \rcp and only a single value is calculated for 
the ratio between 0-40\% to 40-86\% centralities, as shown in 
Figure~\ref{fig:rcp39} and in Table~\ref{tab:39_62_rcp}. The published 
\auau results at 200 GeV are rebinned to have a peripheral centrality 
selection of 40--93\% to approximately match the number of binary 
collisions for the peripheral denominator. Within uncertainties the 
results agree; however, the limited statistics in the \auau at 39 GeV 
preclude any strong conclusions.

The centrality dependence as quantified via \rcp is not a replacement for 
the nuclear modification factor \raa (relative to the \pp baseline) since 
\jpsi yields may change already in peripheral \auau collisions, in 
particular from cold nuclear matter effects.  In addition, \rcp has 
significant uncertainties from the more limited statistics and the larger 
systematic uncertainty on $\langle \ncoll \rangle$ for the peripheral bin.  
The PHENIX experiment has no data for \pp collisions at 39 GeV, and only a 
limited data set was recorded during 2006 for \pp collisions at 62.4 GeV.  
However, \pp measurements do exist from fixed target $p$+$A$ experiments 
near 39 GeV and from ISR collider experiments at 62 GeV. In the Appendix, 
we discuss in detail these results and compare them with theoretical 
calculations within the Color Evaporation Model (CEM) from R. 
Vogt~\cite{Frawley:2008, Vogt:jpsi_cem} to determine a \pp reference.

\begin{figure}
\centering
\includegraphics[width=1.0\linewidth]{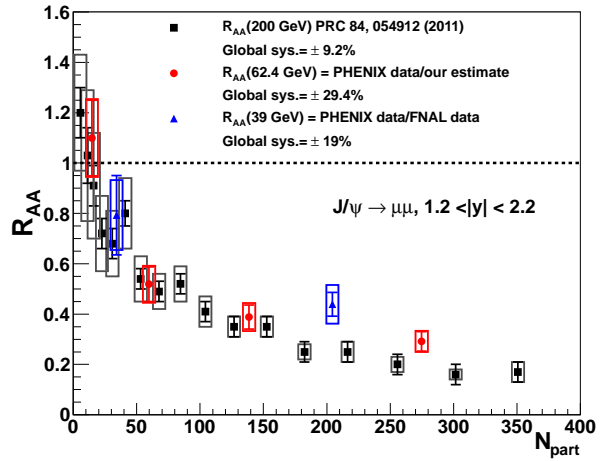}
\caption{\label{fig:raa} (color online) 
\jpsi \raa at \snn = 39, 62.4, and 200 GeV. The solid error bars are the 
quadrature sum of the statistical and type A systematic uncertainties, and 
the boxes represent the correlated (type B) systematic uncertainties. The 
global systematic uncertainties are quoted in the legend for each energy's 
results.
}
\end{figure}

We quantify the nuclear modification factor \raa with respect to the \pp 
reference as follows:

\begin{eqnarray}
\label{equ:raa}
\raa = \frac{1}{\langle\taa\rangle}\frac{dN^{\rm AA}/dy}{d\sigma^{pp}/dy}
\end{eqnarray}

\noindent where $dN^{\rm AA}/dy$ is the invariant yield in \auau 
collisions, $d\sigma^{pp}/dy$ is the \pp cross section, and 
$\langle\taa\rangle$ is the nuclear overlap function (where 
$\langle\taa\rangle=\langle\ncoll\rangle/\sigma_{NN}^{inelastic}$). Unlike 
200 GeV, the 39 and 62 GeV \pp references are determined from other 
measurements rather than being from our own, and systematic uncertainties 
will not cancel in the ratio. Our estimates for the \jpsi \pp cross 
sections in the range $1.2 < |y| < 2.2$ for 39 and 62.4 GeV are shown in 
Table~\ref{tab:39_62_pp}, and are detailed in the Appendix. 
The \jpsi \raa for \auau collisions at 39 and 62.4 GeV is tabulated in 
Table~\ref{tab:39_62_raa} and shown in Figure~\ref{fig:raa} as a function 
of the number of participating nucleons $\langle \npart \rangle$, along 
with the previously published 200 GeV results~\cite{PhysRevC.84.054912}. 
The type C global scale uncertainties, from the \pp references, are listed 
separately in the legend. At both 39 and 62.4 GeV, there is slightly less 
\jpsi suppression than observed in \auau at 200 GeV.  However, 
particularly for 62.4 GeV, since we have no reliable \pp reference from 
our own measurements, the \raa result could shift down by the quoted 29\% 
systematic uncertainty, bringing the data into agreement with the 200 GeV 
result.

\begin{table}[!ht]
\caption{\label{tab:39_62_pp} 
Estimates used for the 39 and 62.4 GeV \jpsi \pp cross sections along with 
their uncertainties.  See the Appendix for details.
}
\begin{ruledtabular} \begin{tabular}{cc}
$\sqrt{s}(GeV)$ & $d\sigma^{pp}/dy$ \\
\hline
39    & 2.91 $\pm$ 19\%~nb\\
62.4  & 7.66 $\pm$ 29.4\%~nb \\
\end{tabular} \end{ruledtabular}
\end{table}

\begin{table}[!ht]
\caption{\label{tab:39_62_raa} 
PHENIX 39 and 62.4 GeV \jpsi $R_{\rm AA}$ vs Centrality with statistical 
uncertainties and Type A, B and C systematics.
}
\begin{ruledtabular} \begin{tabular}{ccccccc}
$\sqrt{s}(GeV)$ & Cent(\%) & $R_{\rm AA}$ & Stat & Type A & Type B & Type C \\
\hline
39       & 0-40  & 0.439 & 0.043 & 0.020 & 0.077 & 0.083 \\
         & 40-86 & 0.793 & 0.157 & 0.011 & 0.139 & 0.151 \\
\\
62.4     & 0-20  & 0.292 & 0.039 & 0.004 & 0.042 & 0.085 \\
         & 20-40 & 0.388 & 0.047 & 0.008 & 0.056 & 0.115 \\
         & 40-60 & 0.519 & 0.067 & 0.014 & 0.073 & 0.153 \\
         & 60-86 & 1.100 & 0.150 & 0.010 & 0.155 & 0.323 \\
\end{tabular} \end{ruledtabular}
\end{table}

\section{Discussion}

\begin{figure}
\centering
\includegraphics[width=1.0\linewidth]{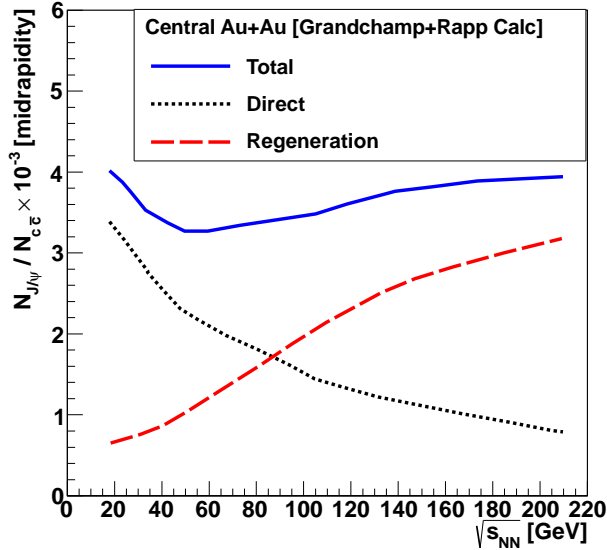}
\caption{\label{fig:rapp} (color online) 
The number of \jpsi per produced charm pair ($\times 10^{-3}$) in \auau 
central collisions ($\npart = 360$) at midrapidity.  Shown are the direct 
\jpsi and regeneration contributions. Calculation details and figure 
from~\cite{Grandchamp:2002wp}.
}
\end{figure}

The collision energy dependence of the various competing effects 
influencing the final \jpsi yields are all quite different. Thus, the 
similarity of the \jpsi nuclear modifications $R_{CP}$ and $R_{\rm AA}$ 
from 39 to 200 GeV is a challenge for models incorporating the many 
effects. There was a prediction that the maximum \jpsi suppression would 
occur near \snn = 50 GeV, as shown in 
Figure~\ref{fig:rapp}~\cite{Grandchamp:2002wp}.  As the collision energy 
increases the QGP temperature increases, and thus the \jpsi color 
screening (labeled as direct \jpsi suppression) becomes more significant. 
However, in this calculation, the regeneration contribution increases with 
collision energy due to the increase in the total number of charm pairs 
produced and nearly compensates. This result is for \jpsi at midrapidity 
and relative to the total charm pair production (thus removing in this 
ratio possible changes in the charm pair production caused by initial 
state effects).

\begin{figure}
\centering
\includegraphics[width=1.0\linewidth]{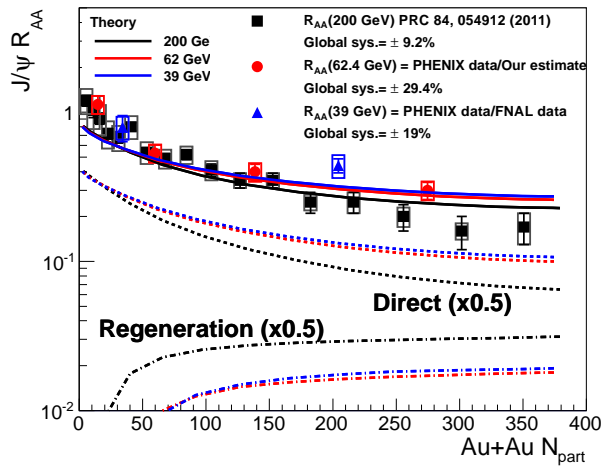}
\caption{\label{fig:rapp_data} (color online) 
The \jpsi nuclear modification factor \raa as a function of the number of 
participating nucleons $N_{\rm part}$ for \snn = 39, 62.4, 200 GeV \auau 
collisions.  Calculation results are shown from~\cite{zhao:2010nk} for the 
total \jpsi \raa and the separate contribution of direct \jpsi suppression 
and regeneration (scaled down by $\times 0.5$ for visual clarity).  The 
PHENIX experimental data points are shown for comparison.
}
\end{figure}

Recently, the same authors have completed new calculations including cold 
nuclear matter effects, regeneration, and QGP suppression specifically for 
\jpsi at forward rapidity~\cite{zhao:2010nk,ZhaoRapp:2010Private}. 
Figure~\ref{fig:rapp_data} shows these results (in the so-called ``strong 
binding scenario'').  The contributions of direct \jpsi and regeneration 
are shown separately (and scaled down by $\times 0.5$ for visual clarity). 
The inclusion of cold nuclear matter effects and the forward-rapidity 
kinematics slightly reverse the trend seen in Figure~\ref{fig:rapp} and 
now the total \jpsi \raa follows the ordering $R_{\rm AA}$(200 GeV) $<$ 
$R_{\rm AA}$(62 GeV) $<$ $R_{\rm AA}$(39 GeV) (though by a very modest 
amount).  Also shown in Figure~\ref{fig:rapp_data} are the PHENIX 
experimental measurements that, within the global systematic 
uncertainties, are consistent with the theoretical calculations.

These results highlight the need for \pp reference data at both 39 and 
62.4 GeV from the same experiment. In addition, the cold nuclear matter 
effects are likely to be different at the different energies (an important 
input for the calculations in Ref.~\cite{zhao:2010nk}).  The $x$ 
distribution of gluons for producing \jpsi at $1.2 < |y| < 2.2$ changes as 
the colliding energy decreases.  In a simple {\sc pythia} study, one finds 
that the average gluon $x_{1}$ and $x_{2}$ for producing \jpsi between 
$1.2 < |y| < 2.2$ is 0.14 and 0.01 for \snn = 200 GeV, 0.32 and 0.03 for 
\snn = 62.4 GeV, and 0.43 and 0.05 for \snn = 39 GeV. The large 
uncertainties in the gluon nPDF for the anti-shadowing and EMC 
regions~\cite{Eskola:2009uj} leads to an additional $\pm 30$\% uncertainty 
in the \jpsi initial production for the central \auau case.  Future 
measurements in $p(d)$+$A$ collisions at these energies are clearly 
required in order to reduce this large uncertainty contribution.

\section{Summary}

The PHENIX experiment has measured the invariant yield of \jpsi at forward 
rapidity in \auau collisions at 39 and 62.4 GeV.  The nuclear 
modification, when formulated as \rcp (the ratio between central and 
peripheral event classes), indicates a similar suppression pattern at the 
two lower energies to that previously published for \auau collisions at 
200 GeV. Using a \pp reference from other experiments and from a Color 
Evaporation Model calculation, results in an \raa with slightly less 
suppression at these lower energies.  These results are consistent with 
theoretical calculations dominated by the balancing effects of more QGP 
suppression as well as more \jpsi regeneration for high-energy collisions. 
However, any firm conclusion regarding the overall level of suppression 
from the QGP requires additional \pp and $p(d)$+$A$ data at these 
energies.

\section{Acknowledgments}


We thank the staff of the Collider-Accelerator and Physics
Departments at Brookhaven National Laboratory and the staff of
the other PHENIX participating institutions for their vital
contributions.  We acknowledge support from the 
Office of Nuclear Physics in the
Office of Science of the Department of Energy, the
National Science Foundation, Abilene Christian University
Research Council, Research Foundation of SUNY, and Dean of the
College of Arts and Sciences, Vanderbilt University (U.S.A),
Ministry of Education, Culture, Sports, Science, and Technology
and the Japan Society for the Promotion of Science (Japan),
Conselho Nacional de Desenvolvimento Cient\'{\i}fico e
Tecnol{\'o}gico and Funda\c c{\~a}o de Amparo {\`a} Pesquisa do
Estado de S{\~a}o Paulo (Brazil),
Natural Science Foundation of China (P.~R.~China),
Ministry of Education, Youth and Sports (Czech Republic),
Centre National de la Recherche Scientifique, Commissariat
{\`a} l'{\'E}nergie Atomique, and Institut National de Physique
Nucl{\'e}aire et de Physique des Particules (France),
Bundesministerium f\"ur Bildung und Forschung, Deutscher
Akademischer Austausch Dienst, and Alexander von Humboldt Stiftung (Germany),
Hungarian National Science Fund, OTKA (Hungary), 
Department of Atomic Energy and Department of Science and Technology (India), 
Israel Science Foundation (Israel), 
National Research Foundation and WCU program of the 
Ministry Education Science and Technology (Korea),
Ministry of Education and Science, Russian Academy of Sciences,
Federal Agency of Atomic Energy (Russia),
VR and Wallenberg Foundation (Sweden), 
the U.S. Civilian Research and Development Foundation for the
Independent States of the Former Soviet Union, 
the Hungarian American Enterprise Scholarship Fund,
and the US-Israel Binational Science Foundation.

\appendix*

\section{Proton-Proton Reference\label{appendixpp}}

\begin{figure}
\centering
\includegraphics[width=1.0\linewidth]{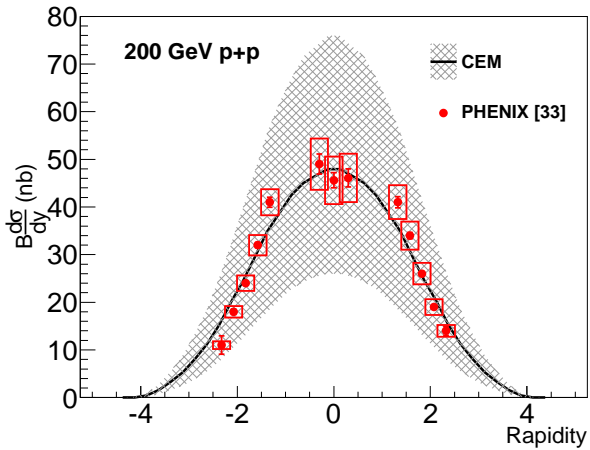}
\caption{\label{fig:ramona200} (color online) 
\jpsi cross section as a function of rapidity in \pp collisions at 200 
GeV.  The CEM calculation is shown as a black solid line with a gray band 
for its uncertainty.  In comparison, PHENIX measurements are shown as red 
points~\cite{Adare:2011vq}.
}
\end{figure}

\begin{figure}
\centering
\includegraphics[width=1.0\linewidth]{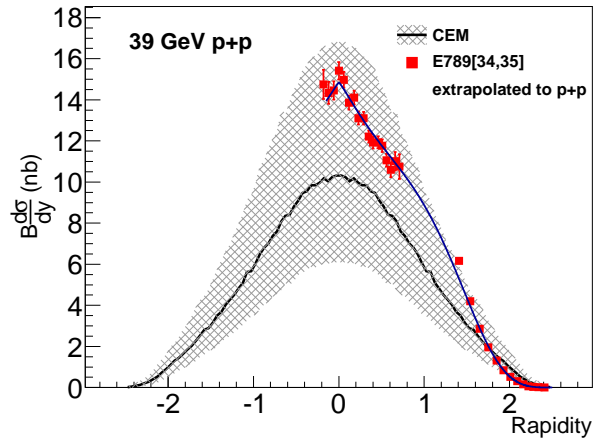}
\caption{\label{fig:ramona39} (color online) 
\jpsi cross section as a function of rapidity in \pp collisions at 39 GeV.  
The CEM calculation is shown as a black solid line with a gray band for 
its uncertainty. Data points and fit function are the result of the 
$p$+$A$ data extrapolation to $p$+$p$ as described in the text.
}
\end{figure}

\begin{figure}
\centering
\includegraphics[width=1.0\linewidth]{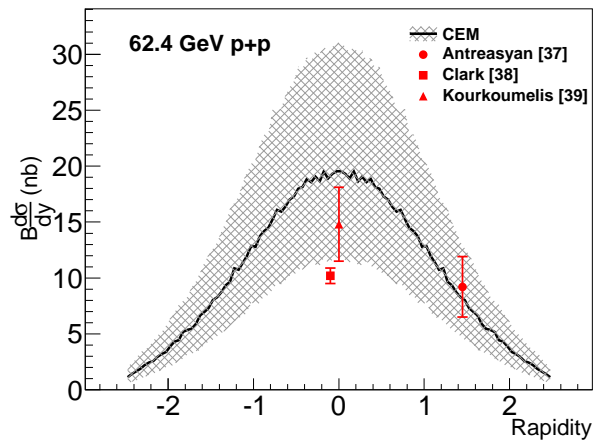}
\caption{\label{fig:ramona62} (color online) 
\jpsi cross section as a function of rapidity in \pp collisions at 62.4 
GeV.  The CEM calculation is shown as a black solid line with a gray band 
for its uncertainty. Data points from ISR measurements are shown as 
detailed in the text.
}
\end{figure}

In order to construct the \pp references at 39 and 62.4 GeV, we utilize 
lower-energy data from Fermilab and the ISR, and also the Color 
Evaporation Model (CEM) calculations from R. Vogt~\cite{Frawley:2008, 
Vogt:jpsi_cem}. These calculations have been extensively compared with 
\jpsi cross sections as a function of center-of-mass energy.  First, shown 
in Figure~\ref{fig:ramona200} is a comparison of the published PHENIX 
measurements for the \jpsi cross section in \pp collisions at 200 
GeV~\cite{Adare:2011vq} and the CEM calculation.  For the CEM calculation, 
the solid line is the central value and the gray band represents the 
systematic uncertainty of the results.  Using the same CEM framework, 
calculation results for \pp at 39 and 62.4 GeV are shown in 
Figs.~\ref{fig:ramona39} and \ref{fig:ramona62}, respectively.  It is 
notable that the predicted cross section at midrapidity drops by 
approximately a factor of 2.5 in going from 200 to 62.4 GeV, and then 
another factor of 1.9 in going from 62.4 to 39 GeV.  The rapidity 
distribution also narrows as expected.

\subsection{\pp at 39 GeV}

Fermilab fixed target experiment E789~\cite{FNAL:E789,FNAL:E789_2} has 
measured the invariant cross sections of \jpsi in $p$+Be, $p$+Cu, and 
$p$+Au collisions over a broad rapidity range at \snn = 38.8 GeV. The 
rapidity coverage for $p$+Au was $-0.1 < y < +0.7$; and for $p$+Be and 
$p$+Cu was $1.4 < y < 2.4$.  In addition, the nuclear dependence of the 
cross sections was measured by E866/NuSea~\cite{FNAL:NuSea} and found to 
follow the functional form, $\sigma_{p+A} = A^{\alpha} \sigma_{p+p}$, 
where,

\begin{equation}
\alpha(x_F)=0.960(1-0.0519x_F-0.338x_F^2)
\end{equation}
(as seen in Figures 2 and 3 of Ref.~\cite{FNAL:NuSea}).  

Using this parametrization for the nuclear dependence, one can extrapolate 
versus $A$ from the $p+$A \jpsi cross sections to those for \pp ($A$ = 1) 
and obtain the \pp cross sections as a function of $x_{F}$. After 
converting these to be versus rapidity, they are shown in 
Figure~\ref{fig:ramona39}. For the rapidity range $1.2 < y < 2.2$ one 
obtains $2.91 \pm 19$\%(syst)~nb by integrating the fit function.  In 
comparison, the result from the CEM calculation is $2.45_{-1.0}^{+1.78}$ 
nb, which agrees well within uncertainties. Thus, we use this extraction 
from the experimental data for the 39 GeV \pp reference, as shown in 
Table~\ref{tab:39_62_pp}. Systematic uncertainties on this reference 
include 12\% from the E789 $p+$A data and 15\% to account for the quality 
of the fit and for its extrapolation in rapidity into the unmeasured $1.2 
< y < 1.4$ region.

\subsection{\pp at 62.4 GeV}

Experiments at the CERN Intersecting Storage Ring (ISR) measured the \jpsi 
cross section in \pp collisions at 62 
GeV~\cite{Antreasyan:1982,Clark:1978mg,Kourkoumelis1980481}. These results 
are shown in Table~\ref{tab:62isr} and in comparison to the CEM 
calculation in Figure~\ref{fig:ramona62}. Since our measurements lie in 
the rapidity range $1.2 < |y| < 2.2$, the most important $p$$+$$p$ 
measurement from the ISR for our purposes is that of 
Antreasyan~\cite{Antreasyan:1982}, which covers a rapidity range of $0.89 
< y < 1.82$ and agrees quite well with the CEM calculation. Therefore we 
estimate the $p$$+$$p$ reference by integrating over our rapidity coverage 
using the CEM calculation fitted to the Antreasyan measurement. For the 
uncertainties of this reference we take similar CEM guided integrals, but 
constrained to the upper and to the lower limits of the that ISR 
measurement. This results in a 62 GeV \pp reference of $7.66 \pm 29.4\%$ 
nb. We note that the midrapidity ISR points are somewhat low but nearly 
consistent with the CEM calculation, but since our data lies at large 
rapidity we rely on the Antreasyan ISR point.

\begin{table}[!ht]
\caption{\label{tab:62isr} 
ISR measurements of \jpsi in \pp collisions at 62 GeV
}
\begin{ruledtabular}\begin{tabular}{ccc}
Reference & Rapidity Range & $B_{ee}\frac{d\sigma}{dy}|$(nb) \\
\hline
Antreasyan {\it et al.}~\cite{Antreasyan:1982} & $0.89 < y < 1.82$ & 9.21 $\pm$ 2.70 \\
Clark {\it et al.}~\cite{Clark:1978mg} & $|y|<0.5$ & 10.2 $\pm$ 0.7 \\
Kourkoumelis {\it et al.}~\cite{Kourkoumelis1980481} & $|y|<0.65$ & 14.8 $\pm$ 3.3 \\
\end{tabular}\end{ruledtabular}
\end{table}



\end{document}